\documentclass[prb,twocolumn,showpacs,floatfix,preprintnumbers,amsmath,amssymb,superscriptaddress]{revtex4-1}
\usepackage{graphicx}
\usepackage{color}

\begin{document}


\title{Evolution of Superconducting Gaps in Th-Substituted Sm$_{1-x}$Th$_x$OFeAs Studied by Multiple Andreev Reflection Spectroscopy}


\author{T.E. Kuzmicheva}
\email[]{kute@sci.lebedev.ru}
\affiliation{P.N. Lebedev Physical Institute, Russian Academy of Sciences, 119991 Moscow, Russia}

\author{S.A. Kuzmichev}
\affiliation{M.V. Lomonosov Moscow State University, Faculty of Physics, 119991 Moscow, Russia}

\author{K.S. Pervakov}
\affiliation{P.N. Lebedev Physical Institute, Russian Academy of Sciences, 119991 Moscow, Russia}

\author{V.M. Pudalov}
\affiliation{P.N. Lebedev Physical Institute, Russian Academy of Sciences, 119991 Moscow, Russia}
\affiliation{National Research University Higher School of Economics}

\author{N.D. Zhigadlo}
\affiliation{Laboratory for Solid State Physics, ETH Zurich, Otto-Stern-Weg 1,
CH-8093 Zurich, Switzerland}
\affiliation{Department of Chemistry and Biochemistry, University of Berne,
Freiestrasse 3, CH-3012 Berne, Switzerland}

\date{\today}

\begin{abstract}
Using intrinsic multiple Andreev reflections effect (IMARE) spectroscopy, we studied SnS contacts in the layered oxypnictide superconductors Sm$_{1-x}$Th$_x$OFeAs with various thorium doping and critical temperatures $T_C = 21\textendash54$\,K. We observe a scaling between both superconducting gaps and $T_C$. The determined BCS-ratio for the large gap $2\Delta_L/k_BT_C = 5.0\textendash5.7$ and its eigen BCS-ratio (in a hypothetical case of zero interband coupling) $2\Delta_L/k_BT_C^L = 4.1\textendash4.6$ both exceeding the weak-coupling limit 3.52, and for the small gap $2\Delta_S/k_BT_C = 1.2 \textendash 1.6$ remain nearly constant within all the $T_C$ range studied. The temperature dependences $\Delta_{L,S}(T)$ agree well with a two-band BCS-like Moskalenko and Suhl model. We prove intraband coupling to be stronger than interband coupling, whereas and Coulomb repulsion constants $\mu^{\ast}$ are finite in Sm-based oxypnictides.
\end{abstract}


\maketitle

\section{Introduction}

Since the discovery of iron-based superconductivity in 2008 \cite{Kamihara}, several families of superconducting ferropnictides were synthesized \cite{Stewart,Johnston}. All iron pnictides possess a layered crystal structure comprising quasi-two-dimensional Fe-As blocks separated by spacers along the $c$-direction. Superconductivity develops namely in Fe-As
layers whose structure remains nearly the same for all the iron pnictides, whereas the difference is in the spacer blocks structure \cite{Stewart,Johnston}. For the so called 1111-family, the whole structure
consists of a stack of superconducting Fe-As blocks and nonsuperconducting $Re$-O spacers ($Re$ is a rare-earth metal).

1111-oxypnictides possess the simplest band structure as compared to other pnictides \cite{Singh}. Band-structure calculations showed that iron $3d$ orbitals make the main contribution to the normal-state density of states (DOS) at the Fermi level, forming electron and hole sheets of the Fermi surface. The hole sheets represent two concentric cylinders near the $\Gamma$ point of the first Brillouin zone, whereas the electron sheets are formed by two cylinders of elliptic cross section near the M points. Both
electron and hole cylinders are slightly warped along the $c$-direction. As was demonstrated in angle-resolved photoemission spectroscopy (ARPES) studies \cite{Charnukha}, these Fermi surface sheets
are considered to be formed by two effective (hole and electron) bands. The ARPES studies \cite{Charnukha} also revealed a feature typical of optimally doped Sm-1111: singular Fermi surface sheets near the $\Gamma$ and M points. Under electron doping, superconducting critical temperatures of SmOFeAs varies in the wide range up to $T_C \approx 57$\,K \cite{Fujioka}. Therefore, Sm-1111 is an ideal candidate for investigating the role of electron doping on the superconducting properties.

To describe multiband superconductivity in iron pnictides, the two basic models were suggested: $s^{++}$-model of coupling through orbital fluctuations enhanced by phonons \cite{Onari1,Onari2}, and $s^{\pm}$-model of spin-fluctuation-mediated superconductivity \cite{MazinRev,Mazin}. To date, both models have not got yet unambiguous experimental evidence. Some theoretical studies predict a certain influence of impurity scattering on the gap values in iron-based superconductors \cite{MMK}. Therefore, direct $\Delta_{L,S}(T_c)$ data are of the most importance to answer the key question concerning the underlying pairing mechanism.

The experimentally determined gap values in Sm-1111 as well as in other oxypnictides in whole are rather contradictory \cite{Naidyuk,Chen,DagheroRu,Daghero,Wang,Noat,Fasano,Millo,Malone}. For example, $2\Delta_L/k_BT_C$ in Sm-1111 determined by point-contact (PCAR) spectroscopy varies by a factor of six, from nearly weak-coupling BCS-limit 3.6 \textendash 3.7 in \cite{Naidyuk,Chen} up to 22 in \cite{DagheroRu}. This fact raises the problem of accurate superconducting order parameter determination by various experimental probes.

Thorium substitution in Sm$_{1-x}$Th$_x$OFeAs oxypnictide supplies charge carriers to superconducting Fe-As layers giving rise to superconductivity. It opens an unique possibility to explore the evolution of the superconducting order parameter versus critical temperature in the same compound with no direct influence to the geometry of Fe-As tetrahedrons \cite{Zhigadlo2010}. To the best of our knowledge, here we present the first data on the evolution of the superconducting gap, $\Delta(T_C)$, and the characteristic ratio, $2\Delta(T_C)/k_BT_C$, for 1111-oxypnictides with heterovalent substitution in wide range of $T_C$. The paper contains a systematic study of current-voltage characteristics (CVC) and dynamic conductance spectra for SnS-Andreev contacts in optimal and underdoped Sm$_{1-x}$Th$_x$OFeAs samples with various thorium doping. Here we present the data in the range $T_C = 35 \textendash 54$\,K and nominal Th concentrations $x = 0.08 \textendash 0.3$, and the pioneer data with $T_C = 21 \textendash 37$\,K sample series ($x \lesssim 0.08$). Using intrinsic multiple Andreev reflections effect (IMARE) spectroscopy, we directly determined the bulk values of two superconducting gaps $\Delta_L$ and $\Delta_S$, their temperature dependences, and BCS-ratios. We found a scaling between both gaps and critical temperature, and nearly constant BCS-ratios within all studied $T_C$ range. We find that the gap temperature dependences $\Delta_{L,S}(T)$ are well described by the two-band Moskalenko and Suhl system of equations \cite{Mosk,Suhl} with a renormalized BCS-integral (RBCS). From this fitting, we have determined the intraband and interband coupling parameters and prove that the intraband coupling is stronger than the interband coupling in Sm-based oxypnictides.

\section{Experimental details}
 \subsection{Synthesis}
Polycrystalline Sm$_{1-x}$Th$_x$OFeAs samples with various thorium doping and critical temperatures ($T_C = 21 \textendash 54$\,K) were
synthesized by high-pressure method. Overall details of the sample cell assembly and high-pressure synthesis process may be found in Refs. \cite{Zhigadlo2010,Zhigadlo2012}. Powders of SmAs, ThAs, Fe$_2$O$_3$, and F of high purity ($\geq 99.95 \%$)
were weighed according to the stoichiometric ratio, thoroughly ground, and pressed into pellets. Then, the pellet containing precursor was enclosed in a boron nitride crucible and placed inside a pyrophyllite cube with a graphite heater. All the preparatory steps were done in a glove box under argon atmosphere. The six tungsten carbide anvils generated pressure on the whole assembly. In a typical run, the sample was compressed to 3\,GPa at room temperature. While keeping the pressure constant, the temperature was ramped up within 1\,h to the maximum value of 1430\,$^{\circ}$C, maintained for 4.5\,h, and finally quenched to the room temperature. Afterward, the pressure was released and the sample removed. Subsequently recorded X-ray powder diffraction patterns revealed high homogeneity of the samples and the presence of a single superconducting phase \cite{Zhigadlo2010}. The amount of additional nonsuperconducting phases SmAs and ThO$_2$ was vanishingly small. The bulk character of superconductivity in Sm$_{1-x}$Th$_x$OFeAs samples was confirmed by magnetization measurements.

\subsection{Preparation of weak links by the break-junction technique}
In our experiments, we used a break-junction technique \cite{Moreland} to produce symmetrical SnS contacts. The sample prepared as a thin rectangular plate with dimensions about $3 \times 1.5 \times 0.1$\,mm$^3$ was attached to a springy sample holder by four-contact pads made of pasty (at room temperature) In-Ga alloy. After cooling down to $T = 4.2$\,K, the sample holder was gently curved, which caused cracking of the bulk sample. The microcrack generates cryogenic clefts and separate the bulk sample into two parts with a weak link between them, thus forming ScS contact (where $c$ is a constriction). Cleavage of a layered sample causes its exfoliation along the $ab$-planes and an appearance of steps and terraces at cryogenic clefts (Fig. 1a). This is typically the case for both single crystals and polycrystalline samples \cite{EPL}. As an illustration, we considered in Ref.~\cite{EPL} a model polycrystalline sample with randomly oriented $ab$-planes of grains, where intergrain connection is just 10 \%
stronger than the interlayer one (along the $c$ direction for any of grains). In this case we expect quite a considerable amount of split crystallites in the ab-plane $(2 \textendash 6\%)$.
In a more realistic situation,  when the strength of intergrain connection exceeds the interlayer ultimate strength by $20\%$,
about $4 \textendash 11\%$
of grains would split, causing the appearance of large amount of steps and terraces. These estimates are supported by the electron microscope image of polycrystalline sample cleft shown in Fig.~1\,b.

\subsection{SnS Andreev junction and arrays of junctions}

Under fine tuning of the sample holder curvature, the two cryogenic clefts slide apart touching through various terraces. This enables
to vary the cross-size of the resulting ScS contact in order to realize a ballistic regime. In the majority of Fe-based superconductors we studied, the constriction is electrically equivalent to a thin layer of normal metal, and the resulting current-voltage characteristic (CVC) and $dI(V)/dV$ are typical for clean classical SnS-Andreev junction with high transparency of about $95 \textendash 98 \%$
\cite{Andreev,OTBK,Arnold,Averin}.
Such contacts exhibit a multiple Andreev reflections effect which manifests itself as a pronounced excess current at low bias voltages (so called foot) in CVC, and a subharmonic gap structure (SGS) in the $dI(V)/dV$ spectrum. At temperatures below $T_C$ SnS-contact demonstrates an excess conductance at any bias, whereas the SGS represents a series of dynamic conductance minima at certain positions:

\begin{equation}
V_n(T) = 2\Delta(T)/en,
\end{equation}

where $n$ is a natural subharmonic order.
In principle, the first Andreev minimum could be slightly shifted towards lower biases, $V_{n_L=1} \lesssim (2\Delta/e)$ \cite{OTBK,Arnold,Kummel,Averin}. If so, the gap value may be determined from
positions of the higher order SGS dips with $n \geqslant 2$. In case of a two-gap superconductor, two subharmonic gap structures should be expected. The number of observed SGS dips strongly depends on the ratio between the carrier mean free path $l$ and the contact size $a$: $n_{max} \approx l/2a$ \cite{Kummel}.

The break-junction experiments with layered samples, beside the single SnS contacts, also show arrays of SnS contacts \cite{EPL}.
In this case, the CVC and $dI(V)/dV$ demonstrate Andreev minima at positions which are integer multiplies $m$ of those for the single SnS junction:

\begin{equation}
V_n(T) = m \times 2\Delta(T)/en.
\end{equation}
This obviously corresponds to a stack of $m$ sequentially connected identical SnS junctions. The numbers $m$ can be easily determined by comparing $dI(V)/dV$ curves for various arrays: after scaling the bias voltage axis by $m$, the positions of SGS dips in dynamic conductance spectra should coincide.

The Andreev dips in CVC and $dI(V)/dV$ for such arrays are more pronounced than those for single SnS junction; the larger $m$, the sharper peculiarities are usually observed \cite{EPL}. This firm experimental fact indicates that the origin of such arrays with high-quality characteristics could not be thought about as a chain of independent nonequivalent grain-grain contacts \cite{remark_1}.
By contrast, probing the Andreev arrays ensures one to minimize surface defects influence and measure namely \textit{bulk properties} of the sample \cite{EPL}. The intrinsic multiple Andreev reflections effect (IMARE) occurring in such arrays is similar to the intrinsic Josephson effect in SIS contacts (where $I$ is an insulator) \cite{Nakamura}; both effects were observed first in cuprates \cite{PonomarevIMARE,PonomarevIJE}.

\begin{figure}
\includegraphics[width=20pc]{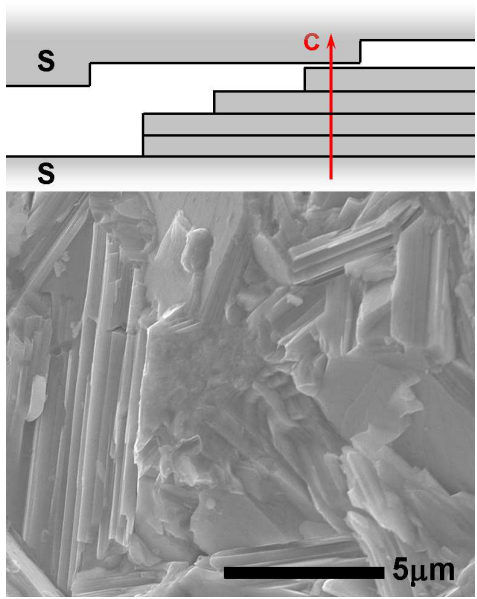}
\caption{a){Schematic drawing} of steps and terraces touching each other in the microcrack a layered sample. The current flowing along the $c$-direction is depicted by arrow. b) The electron microscope image of cleft in Sm$_{1-x}$Th$_x$OFeAs demonstrating steps and terraces at the surface of cracked crystal grains.}
\end{figure}

The IMARE spectroscopy realized by the break-junction technique has a number of advantages:

a) the microcrack generates terraces of  about atomic size.
They remain tightly connected during sliding that prevents impurity penetration into the microcrack and protects the purity of cryogenic clefts;

b) the contact point is far from current and potential leads, which prevents junction overheating and provides true four-point connection;

c) by fine bending of the sample holder, one can probe several tens of Andreev arrays with various diameter and number of junctions in the stack $m$ in one and the same sample and during the same cooldown; it enables to collect statistics and to check the data reproducibility;

d) unlike asymmetric NS and NIS junctions \cite{BTK,Dynes}, in SnS-Andreev contacts the gap value may be determined directly (from the positions of SGS dips), and no fitting of $dI(V)/dV$ is needed; the latter remains true at any temperatures $0 \leq T < T_C$ \cite{OTBK,Kummel}, therefore, one can obtain precise temperature dependences of the gaps;

e) by probing the Andreev arrays one unambiguously determines bulk values of superconducting gaps.

The dynamic conductance spectra were measured directly by a standard modulation technique \cite{LOFA}. We used a current source with ac frequency less than 1\,kHz. The results obtained with this setup are insensitive to the presence of parallel ohmic conduction paths; if any path is present, dynamic conductance curves shift along the vertical axis only, while the bias stay unchanged.

\section{Results and discussion}
\subsection{IMARE in optimally doped samples}

\begin{figure}
\includegraphics[width=20pc]{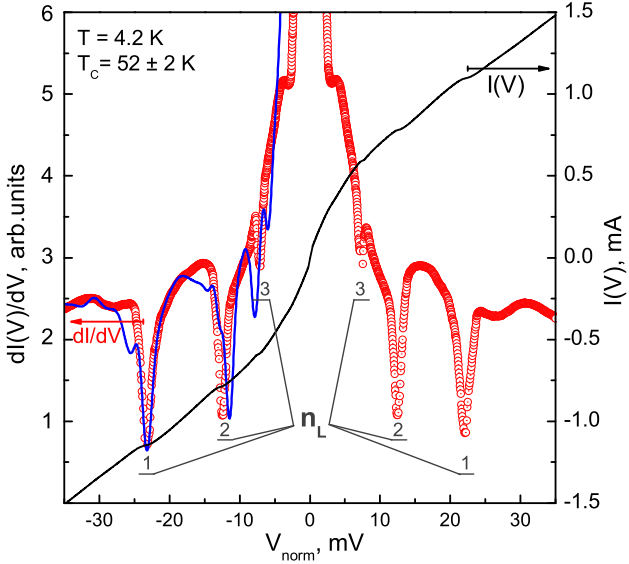}
\caption{Dynamic conductance spectrum (circles, left scale) and current-voltage characteristic (black line, right scale) measured at $T = 4.2$\,K for SnS Andreev contact in optimal Sm-1111 sample with critical temperature $T_C^{bulk} = 52 \pm 2$\,K and nominal $x \approx 0.3$. Blue line corresponds to a rough dI(V)/dV fit based on the MARE model \cite{Kummel}. Gray lines and $n_L$ label indicate the subharmonic gap structure dips for the large gap $\Delta_L \approx 11.9$\,meV. The bias voltage is normalized to that for a single contact.}
\end{figure}

Figure 2 shows normalized CVC (black line; right Y-axis) and dynamic conductance (red line; left Y-axis) for ScS array formed at $T = 4.2$\,K in nearly optimal Sm-1111 sample ($\sharp 2$) with critical temperature $T_C = 52 \pm 2$\,K and nominal thorium concentration $x \approx 0.3$. The array contains $m = 3$ ScS junctions; in order to normalize CVC and $dI(V)/dV$ to those for a single junction, the X-axis was divided by a factor of 3 in Fig.~2. The CVC has a pronounced foot area at low bias voltages. The excess current there is larger than that in NS-contact, where the low-bias conductance is about twice larger than at high-bias \cite{BTK}. The CVC and dynamic conductance spectrum are typical for a highly-transparent ($\approx 95 \%$)
SnS-Andreev contact \cite{Kummel,Averin}. Obviously, the theoretical dependence (blue curve in Fig. 2) based on the MARE model \cite{Kummel} extended for the case of $\thicksim 10\,\%$ gap anisotropy
fits the experimental data (circles) very well. The model \cite{Kummel}, beside $l/2a$ ratio, accounts finite temperatures and possible presence of an Andreev band within the gap. The latter causes the complex fine structure in the fit (satellite dips beyond the subharmonics) unobservable in the experiment; this issue requires a special study. A slight deviation from the expected position (formula (1)) of Andreev dips ($10 \,\%$ uncertainty) is rather conventional. For details, see the Appendix.
Since the four subharmonics are observable (the $n=4$ feature is resolved in $d^2I/dV^2$, not shown here), the effective contact diameter is less though comparable to the mean free path, $l/2a \approx 3 \textendash 4$. This is the reason why the intensity of SGS dips in the experimental spectrum decreases more rapidly as compared to the fit, where $l/2a = 5$. Nevertheless, the clear SGS is the strong evidence for MARE realized in ballistic SnS-contact only.

The other way to check whether the contact is ballistic, is to take a normal-state bulk resistivity for optimal Sm(Th)-1111 single crystal $\rho \approx 0.09$\,m$\Omega \cdot \rm{cm}$\, \cite{Zhigadlo2010}, the average product of bulk resistivity and carrier mean free path $\rho l^{el} \approx 5 \times 10^{-10}$\,$\Omega \cdot \rm{cm^2}$ for Sm-1111 \cite{Tropeano1,Tropeano2} which implied to be nearly constant. The resulting elastic mean free path value $l^{el} \approx 55$\,nm for our sample. Then, taking the resistance of single ScS junction in the array under study $R \approx 25$\,$\Omega$ (see Fig.2), and using Sharvin's formula for a ballistic ($a < l$) contact \cite{Sharvin}: $R = \frac{4}{3\pi}\frac{\rho l}{a^2}$, we get $a \approx 28~{\rm nm} < l^{el}$, thus proving the contact to be ballistic. Going into details, for the experimental observation of MARE namely $l^{in}/2a$ ratio is essential ($l^{in}$ \textemdash inelastic mean free path). Usually, $l^{in}$ is several times larger than $l^{el}$ well providing the ballistic regime. The estimated contact diameter is also many times smaller than the typical crystallite dimension $\thicksim 70 \times 70 \times 20$\,$\mu \rm{m^3}$ \cite{Zhigadlo2010}. The latter confirms the assumption that the SnS array was formed on steps and terraces of a split crystallite.

\begin{figure}
\includegraphics[width=20pc]{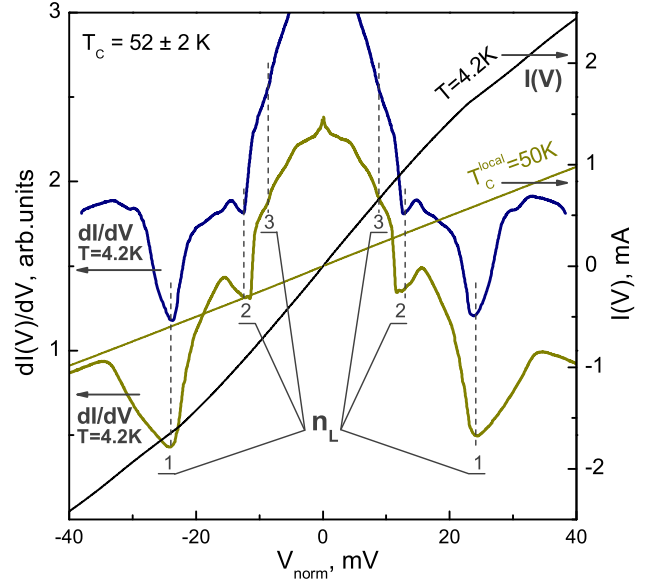}
\caption{Normalized dynamic conductance spectra (left scale) measured at $T = 4.2$\,K for SnS Andreev contacts in optimal Sm-1111 samples with critical temperatures $T_C^{bulk} = 52 \pm 2$\,K. The number of SnS junctions in the arrays are $m = 7$ (upper spectrum), and $m = 2$ (bottom spectrum). Gray lines and $n_L$ label indicate the subharmonic gap structure dips for the large gap $\Delta_L = 12.3 \pm 1.2$\,meV. $I(V)$ characteristics (right scale) corresponding to the bottom $dI(V)/dV$ curve measured at $T = 4.2$\,K and at $T = T_C^{local} \approx 50$\,K are shown for comparison.}
\end{figure}

The numbers and their underlining horizontal strips in Fig.~2 mark the positions and error bars of sharp dips in the dynamic conductance located at $|V_{n_L=1}| \approx 23.5$\,mV, $|V_{n_L=2}| \approx 12.4$\,mV, and $|V_{n_L=3}| \approx 7.6$\,mV. These figures satisfy Eq.~(1) as the first, second and third SGS dip for the large gap $\Delta_L \approx 11.9$\,meV with the BCS-ratio $2\Delta_L/k_BT_C \approx 5.3$. The SGS minima have similar shape and become less intensive with subharmonic order $n$ increasing; this is in accord with theory \cite{Kummel}. The interpretation of the minima in Fig.~2 is straightforward. For example, the minima at $\approx 23.5$\,mV and $\approx 12.4$\,mV cannot be considered as $n=2$ and $n=3$ SGS harmonics, respectively. As follows from Eq.~(1), the ratio $V_n/V_{n+1} = 2$ is true only when $n=1$. Weaker peculiarities at $|V| \approx 2.9$\,mV are located neither at the expected positions of the fourth SGS dips $|V_{n_L=4}| \approx 6.2$\,mV, nor of the small gap $|V_{n_S=1}| \approx 6$\,mV (as was shown in our previous studies \cite{SmJETPL} of Sm-1111; this gap is not identified reliably and might be interpreted as the beginning of the foot area.

\begin{figure}
\includegraphics[width=20pc]{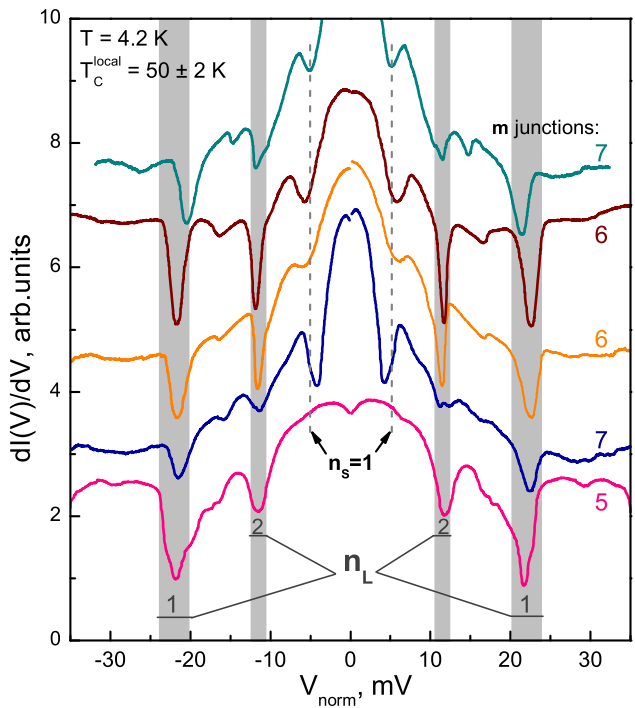}
\caption{Normalized dynamic conductance spectra measured at $T = 4.2$\,K for Andreev arrays in optimal Sm-1111 samples with critical temperatures $T_C = 50 \pm 2$\,K. The number of SnS-junctions in the arrays (from the top) are $m = 7, 6, 6, 7, 5$, correspondingly. Gray vertical areas and $n_L$ label indicate subharmonic gap structure dips for the large gap $\Delta_L = 11.5 \pm 1.2$\,meV. Vertical dashed lines, arrows and $n_S$ label point to the Andreev peculiarities for the small gap $\Delta_S = 2.5 \pm 0.5$\,meV.}
\end{figure}

All the contact properties described above (the presence of the foot area and excess conductance, SGS and ballistic regime) prove these break-junctions to be namely SnS-junctions with high-transparent interface. The same is true for the experimental data presented below.

\begin{figure}
\includegraphics[width=20pc]{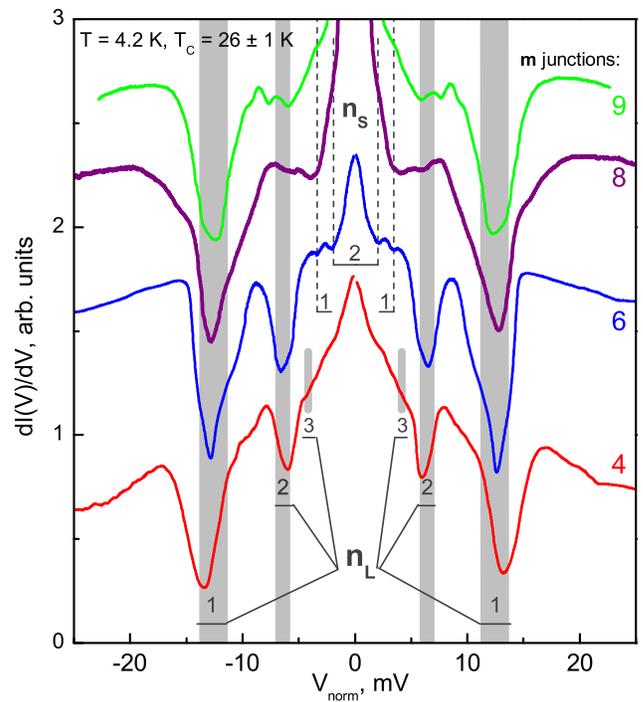}
\caption{Normalized dynamic conductance spectra measured at $T = 4.2$\,K for Andreev arrays in underdoped Sm-1111 samples with critical temperatures $T_C^{bulk} = 26 \pm 1$\,K. The number of SnS-junctions in the arrays (from the top) are $m = 9, 8, 6, 4$, correspondingly. The $n_L$ label and gray vertical areas indicate subharmonic gap structure dips for the large gap $\Delta_L = 6.3 \pm 1.0$\,meV. Dashed vertical lines and $n_S$ label point to the Andreev peculiarities for the small gap $\Delta_S = 1.7 \pm 0.3$\,meV.}
\end{figure}

The normalized dynamic conductance spectra measured at $T = 4.2$\,K for SnS-arrays of $m = 2$ (lower curve) and $m = 7$ (upper curve) junctions in the stacks are compared in Fig.~3. The data were obtained in different nearly optimal Sm-1111 samples with the same critical temperature $T_C = 52 \pm 2$\,K. The $dI(V)/dV$ curves were offset vertically for clarity. For the two-junction array (obtained in sample $\sharp 18$), we also show CVC with excess current measured at $T = 4.2$\,K and linear CVC measured close to the local critical temperature $T_C \approx 50$\,K (corresponding to the transition of the contact area with dimension $\thicksim 10$ -- 30\,nm to the normal state).

The contact resistance increases with temperature, from $R(4.2\,{\rm K}) \approx 16$\,$\Omega$ to $R(50\,{\rm K}) \approx 41$\,$\Omega$ which agrees with the theory predictions for ballistic SnS-contacts \cite{Klapwijk}. The dynamic conductance spectra demonstrate pronounced dips at $|V_{n_L=1}| \approx 24$\,mV, $|V_{n_L=2}| \approx 12.3$\,mV being the SGS minima of $n=1,2$ order. As for these contacts $a \approx 35 {\rm nm} \approx 0.6 l$, the third-order Andreev peculiarities at $|V_{n_L=3}| \approx 8.3$\,mV are strongly smeared. Remarkably, despite the $dI(V)/dV$ in Fig.~3 were obtained with different samples, the dynamic conductance spectra look very similar. The resulting gap value $\Delta_L \approx 12.3$\,meV with $2\Delta_L/k_BT_C \approx 5.5$ is reproducible for both samples.

If we assume that the lower $dI(V)/dV$ is produced by $m = 3$ rather than by $m = 2$ junction array, we immediately obtain the large gap value $\Delta_L \approx 8$\,meV leading to $2\Delta_L/k_BT_C \approx 3.6$ which seems to be too low for 1111 pnictides \cite{EPL,UFN,LOFA,PonFPS,SmJETPL}. For another SnS array presented in Fig.~3 (upper $dI(V)/dV$, sample $\sharp 3$), the bias voltage of its raw dynamic conductance was divided by $m = 7$. After such normalization, the positions of the main gap peculiarities are in good agreement, thus demonstrating IMARE for Sm-1111. Herewith, the dynamic conductance of 7-junction array shows sharper Andreev dips than those of the 2-junction array. This could be due to diminishing of surface influence on superconducting properties of arrays with a large $m$ \cite{EPL}.

\begin{figure}
\includegraphics[width=20pc]{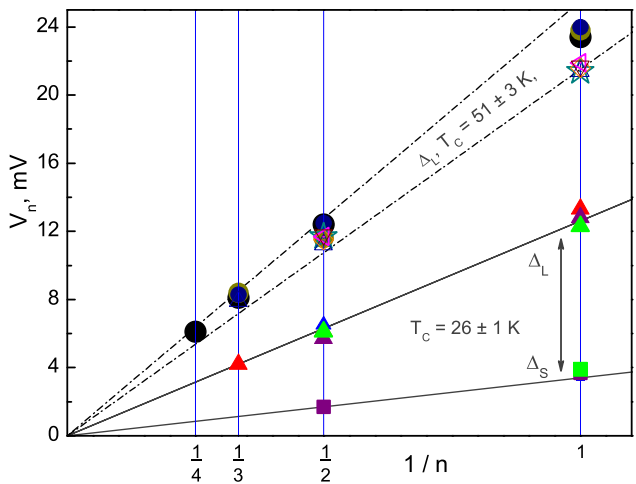}
\caption{The positions of subharmonic gap structure dips $V_n$ versus the inverse number $1/n$. The $V_n$ of the large gap for Andreev contacts with maximal $T_C = 52 \pm 2$\,K (see Figs. 2,3) are shown by solid circles, the data related to Sm-1111 with $T_C = 50 \pm 2$\,K (see Fig. 4) are shown by open symbols. For underdoped Sm-1111 with $T_C = 26 \pm 1$\,K (see Fig. 5), the data for the large gap are shown by solid triangles, for the small gap \textemdash by solid squares of corresponding color. Gray lines are guidelines.}
\end{figure}

A number of dynamic conductance spectra measured at $T = 4.2$\,K for Andreev arrays with various number of junctions $m$ obtained in nearly optimal samples with $T_C = 50 \pm 2$\,K are presented in Fig.~4. The large gap minima are marked with gray vertical areas and  $n_L = 1,2$ labels. The position of the first SGS minimum is slightly shifted from the expected $|V_{n_L=1}| = 2\Delta_L/e$ position \cite{Kummel}, therefore it is reasonable to determine the large gap value from the second SGS dip. Four upper curves with $m = 6, 7$ were obtained with one and the same sample $\sharp 3$ by a fine mechanical tuning. Under the gentle readjustment, the number of SnS-junctions in the stack varied by one, therefore, in the raw $dI(V)/dV$ characteristics the position of the second Andreev dip jumped by $\pm \Delta/e$. Taking the difference between $n_L = 2$ positions, we normalized the spectra by corresponding natural numbers $m$ and got the large order parameter $\Delta_L \approx 11.5$\,meV with $2\Delta_L/k_BT_C \approx 5.3$. We stress again good reproducibility of the spectra and their fine structure.

The lower curve in Fig.~4  obtained with another sample ($\sharp 1$) corresponds to a 5-junction array. At lower biases, in each spectra one can see features at $|V_{n_S=1}| \approx 5$\,meV, which we interpret as the main Andreev peculiarities for the small gap $\Delta_S \approx 2.5$\,meV ($2\Delta_S/k_BT_C \approx 1.2$). Note that the latter bias voltages do not coincide with the expected $|V_{n_L=4}| \approx 5.8$\,mV for the fourth-order $\Delta_L$ peculiarities. Analyzing our data on nearly optimal Sm-1111, we note that the small gap peculiarities are observed not in each spectra. One may suggest several reasons for the strongly smeared SGS of the small gap, including small mean free path in the bands with $\Delta_S$. The specific band structure in Sm-1111 also may contribute: as revealed by recent ARPES studies \cite{Charnukha}, the respective FS sheets are not cylinders and have singularities in optimal Sm-1111. Nonetheless, the positions of peculiarities marked as $n_S=1$ are scaled by $m$, the resulting $\Delta_S$ value and temperature dependence $\Delta_S(T)$ are reproducible, thus showing the bulk nature of these peculiarities.

\begin{figure}
\includegraphics[width=20pc]{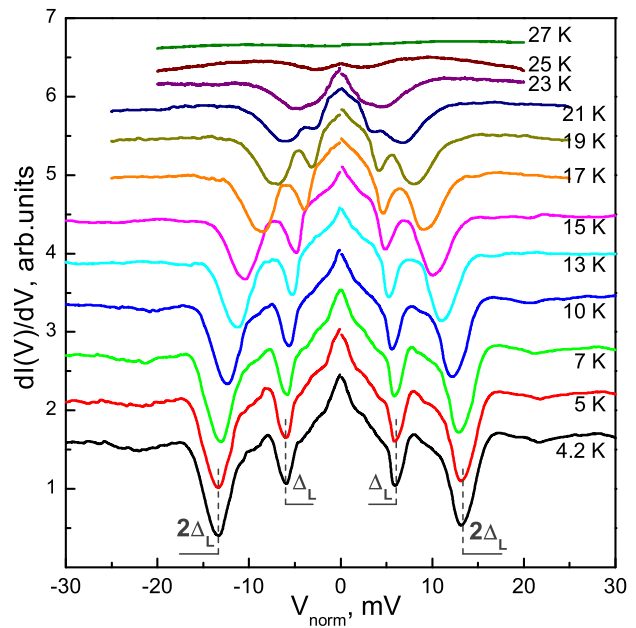}
\caption{Normalized dynamic conductance spectrum in underdoped Sm-1111 (see Fig. 5, lower curve) measured at $T = 4.2$ -- 27\,K. The spectra were offset vertically for clarity, nevertheless, the contact conductance decreases with temperature. Linear background was suppressed. Local critical temperature is $T_C^{local} = 26 \pm 1$\,K. The gray vertical dashed lines indicate subharmonic gap structure dips ($n_L = 1,2$) for the large gap $\Delta_L \approx 6.3$\,meV. The $n_L=3$ subharmonic is poorly visible.}
\end{figure}

\subsection{Underdoped samples}
We also observed IMARE with underdoped Sm-1111 samples with nominal thorium concentration $x \lesssim 0.08$. Figure 5 shows excess-conductance $dI(V)/dV$ curves for Andreev arrays formed at $T = 4.2$\,K in the samples with a factor of two lower critical temperature, $T_C = 26 \pm 1$\,K. The array (presented by the upper dynamic conductance spectrum in Fig.~5) was obtained in sample $\sharp 24$, whereas three other curves correspond to SnS arrays formed in another sample $\sharp 21$. Selecting natural numbers $m = 9, 8, 6, 4$, we achieve a coincidence between the positions of the large gap SGS (marked as $n_L = 1, 2, 3$ and highlighted by gray vertical areas in Fig.~5), and for the small gap peculiarities (dashed lines, $n_S = 1, 2$ label). Intensive minima of the first and second order located at $|V_{n_L=1}| \approx 12.6$\,mV, $|V_{n_L=2}| \approx 6.3$\,mV, and third-order peculiarities at $|V_{n_L=3}| \approx 4.2$\,mV unambiguously determine the large gap $\Delta_L \approx 6.3$\,meV. For the highest quality Andreev spectra in underdoped Sm-1111 (see dynamic conductance for the 6-junction stack in Fig.~5), we also observe SGS for the small gap comprising the first ($|V_{n_S=1}| \approx 3.3$\,mV) and the second ($|V_{n_S=2}| \approx 1.7$\,mV) peculiarities. This gives the small gap value $\Delta_S \approx 1.7$\,meV. The determined values of both gaps are reproducible. $dI(V)/dV$ curves are symmetrical and have no signatures of overheating.

\begin{figure}
\includegraphics[width=20pc]{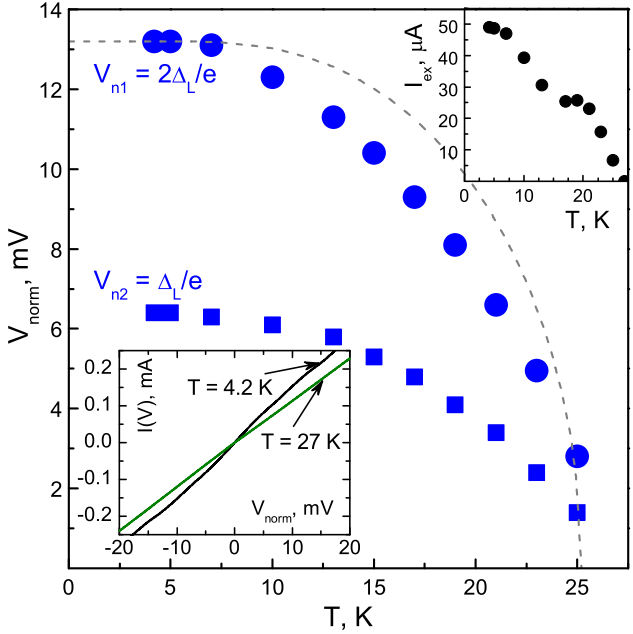}
\caption{Temperature dependence of the positions of the first (circles) and the second (squares) Andreev dips for the large gap in the $dI(V)/dV$ shown in Fig. 7. The upper inset shows the temperature dependence of excess current in $I(V)$ for this contact. The lower inset shows the change in current-voltage characteristic at $T = 4.2$\,K, and at $T = 27$\,K.}
\end{figure}

A summary of the data for SnS contacts obtained in nearly optimal and underdoped Sm-1111 samples is presented in Fig.~6. According to Eqs.~(1,2) positions $V_n$ of the SGS dips should depend linearly on their inverse order $1/n$, and the line should also pass the origin. The $V_n$ positions of the large gap peculiarities for optimal samples with $T_C =52 \pm 2$\,K (see Figs.2,3) are shown by solid circles, for the samples with $T_C = 50 \pm 2$\,K (see Fig.~4) \textemdash by open symbols. The experimental points are confined into the segment (dash-dot lines) passing through the (0;0)-point; the $V_n$ dispersion is obviously caused by the $T_C$ variation. The average gap values are: $\Delta_L = 12.4 \pm 1.2$\,meV for Sm-1111 with $T_C \approx 52$\,K, $\Delta_L = 11.5 \pm 1.2$\,meV, $\Delta_S = 2.5 \pm 0.5$\,meV for Sm-1111 with $T_C \approx 50$\,K. For underdoped samples with $T_C = 26 \pm 1$\,K, the large gap SGS positions are presented in Fig.~6 by triangles, the SGS positions for $\Delta_S$ \textemdash by squares. The data demonstrate two linear dependences starting from the origin. For $T_C \approx 26$\,K, we get average values $\Delta_L = 6.3 \pm 0.6$\,meV and $\Delta_S = 1.7 \pm 0.3$\,meV. The corresponding BCS-ratios $2\Delta_L/k_BT_C \approx 5.6$, $2\Delta_S/k_BT_C \approx 1.5$ are nearly the same as obtained for Sm-1111 with high $T_C$.


\begin{figure}
\includegraphics[width=20pc]{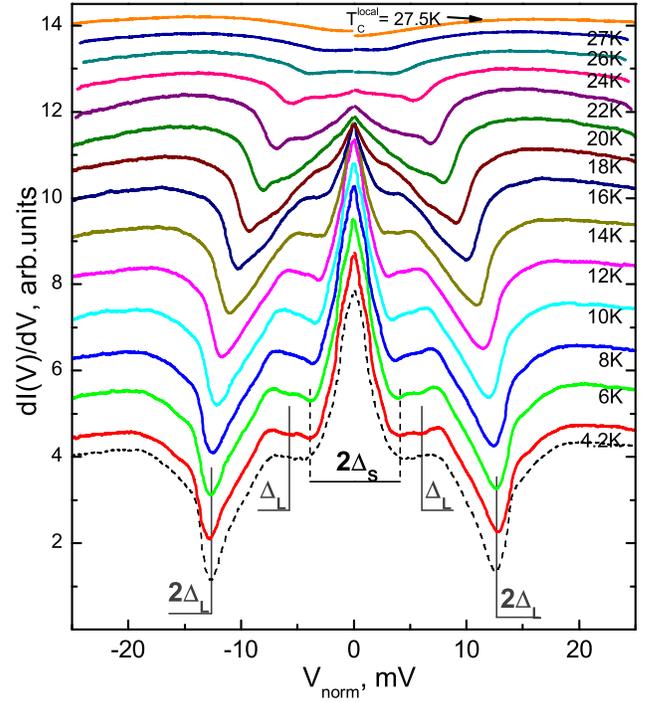}
\caption{Normalized dynamic conductance spectra measured at $T = 4.2$ \textendash 27.5\,K for Andreev array (2$^{nd}$ curve from the top in Fig. 5) in underdoped Sm-1111. The spectra were offset vertically for clarity, nevertheless, the contact conductance decreases with temperature. Local critical temperature is $T_C^{local} = 27.5 \pm 1$\,K. The $n_L$ labels and vertical lines indicate subharmonic gap structure for the large gap $\Delta_L \approx 6.3$\,meV. Dashed lines and $2\Delta_S$ label point to the SGS for the small gap $\Delta_L \approx 2.0$\,meV. Lower dashed spectra ($T=4.2$\,K) was recorded after the thermal cycling, to demonstrate the mechanical stability of the break-junction.}
\end{figure}

\subsection{Temperature dependence of the superconducting gaps}
Temperature evolution of the dynamic conductance spectrum of Andreev array in underdoped Sm-1111 sample (see Fig. 5, lower curve) is shown in Fig.~7. The $dI(V)/dV$ curves are offset with temperature increase, and the linear background is subtracted. The lower spectrum (measured at $T = 4.2$\,K) in Fig. 7 demonstrates clear SGS for the large gap (the positions of the first and the second SGS dips are labeled as $2\Delta_L$ and $\Delta_L$, respectively). As  temperature increases, the dips move towards zero bias, whereas the upper spectrum (measured at $T = 27$\,K) in Fig.~7 becomes nearly linear which corresponds to the normal state. Similarly to the Andreev arrays in nearly optimal samples (see Fig.~3), the contact resistance increases with the temperature, as shown in the lower inset of Fig.~8. At $T = 4.2$\,K, the contact resistance is $R \approx 70$\,$\rm{\Omega}$ and is large enough to provide a ballistic mode. The excess current probed at high bias voltage $eV \approx 2\Delta_L(4.2\rm{K})$ being maximal at $T = 4.2$\,K turns to zero at $T_C^{local}$ as shown in the upper inset of Fig.~8.

Positions of the first (circles) and the second (squares) dynamic conductance minima versus temperature, corresponding to $2\Delta_L(T)$ and $\Delta_L(T)$ dependences \cite{Kummel}, are presented in Fig.~8. Both $dI(V)/dV$ peculiarities have similar temperature dependence, thus proving these peculiarities to be related to the same SGS. The dependences deviate from the single-gap BCS-like curve (dashed line in Fig.~8) being slightly bent down in comparison with the BCS-type $T$-dependence. Since the data of Fig. 8 are obtained for SnS-array and demonstrate namely bulk properties, it does not represent the surface gap. Thus, the observed deviation of the temperature course points to the presence of the second superconducting condensate and the respective gap $\Delta_S$, which was not resolved by IMARE spectroscopy (see Fig.~7). The latter could be due to a low concentration of carriers in the bands with $\Delta_S$ \cite{Charnukha}.

\begin{figure}
\includegraphics[width=20pc]{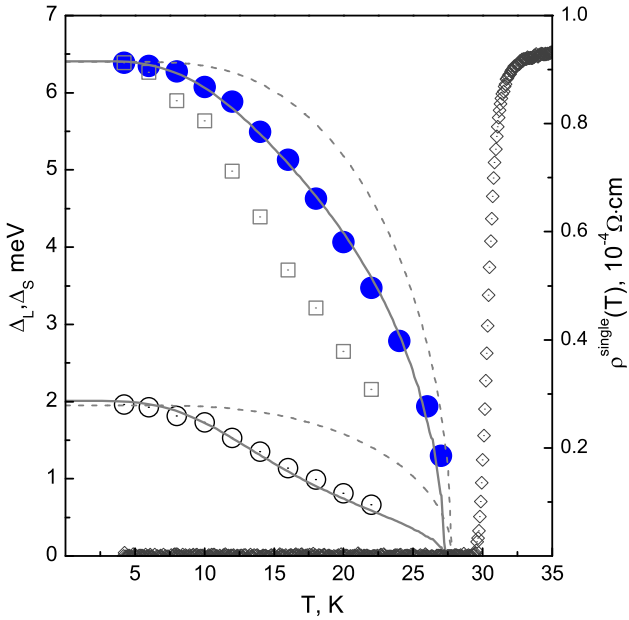}
\caption{Temperature dependence of the large gap (solid circles) and of the small gap (open black circles) for underdoped Sm-1111 (see Fig. 9). Local critical temperature is $T_C^{local} = 27.5 \pm 1$\,K. The normalized dependence $\Delta_S(T)/\Delta_S(0) \times \Delta_L(0)$ is presented by squares for comparison. Theoretical fit by two-gap Moskalenko and Suhl equations \cite{Mosk,Suhl} is shown by solid lines, single-gap BCS-like curves are shown by dashed lines. Bulk resistive transition (for a sample with nominal $x < 0.08$) is shown by open rhombs (right scale).}
\end{figure}

\begin{figure}
\includegraphics[width=20pc]{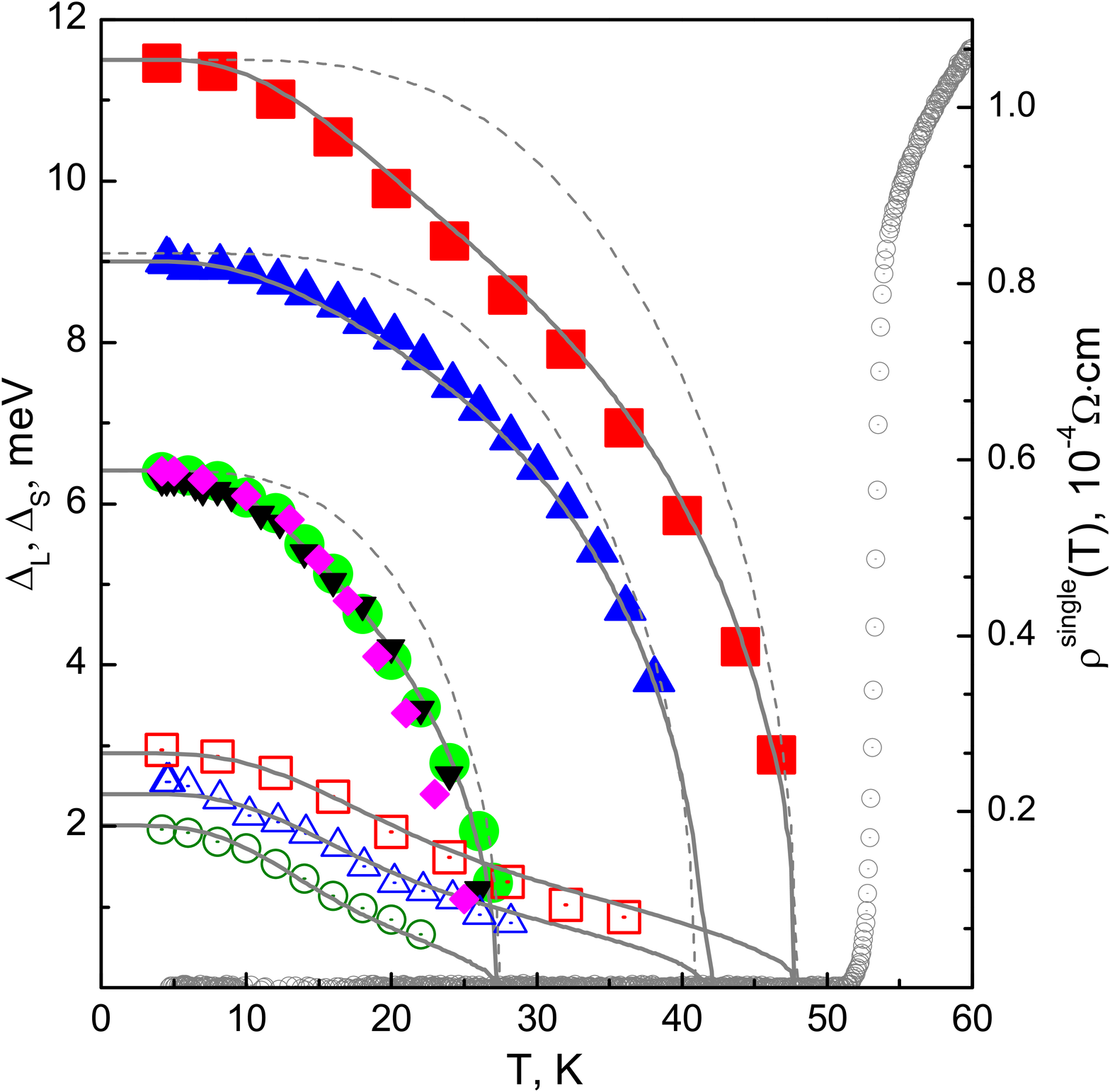}
\caption{Temperature dependences of the large gap (solid symbols) and of the small gap (open symbols of corresponding colour and shape) for Sm-1111 samples with various thorium doping. $T_C^{local} = 26 \textendash 49$\,K. The $\Delta(T)$ shown by blue circles are similar to those in Fig. 10. Theoretical fits by two-gap Moskalenko and Suhl equations \cite{Mosk,Suhl} are shown by solid lines, single-gap BCS-like curves are shown by dashed lines. Temperature dependence of bulk resistivity near superconducting transition ($x \approx 0.3$) is shown by gray open circles (right scale).}
\end{figure}

Figure 9 shows temperature evolution of the dynamic conductance for another Andreev array measured with the same sample as that shown in Fig.~7. Here, the features of the weaker condensate are more clearly pronounced. At $T = 4.2$\,K, the SGS peculiarities for the large gap $\Delta_L \approx 6.3$\,meV are labelled as $2\Delta_L$ and $\Delta_L$; the position of the first peculiarity for the small gap $\Delta_S \approx 2$\,meV is labelled by $2\Delta_S$. The spectra are offset vertically with temperature increase.

The dashed-line spectrum corresponds to $dI(V)/dV$ measured at liquid helium temperature after thermocycling (to $T_C$ and back). The spectrum remains quantitatively similar to the initial $dI(V)/dV$ measured at $T = 4.2$\,K. The reproducibility of the spectra demonstrates high mechanical stability of the break junction. The positions of both SGS peculiarities decrease with temperature and turn to zero at local critical temperature $T_C^{local} \approx 27.5$\,K. The temperature dependences for the large gap (solid circles) and for the small gap (large open circles) were directly determined similarly to Fig.~8; they are  presented in Fig. 10. The $\Delta_L(T)$ temperature dependence slightly bends down as compared to the BCS-type curve shown by the dashed line. As temperature increases, the small gap starts decreasing more rapidly, then almost linearly tends to the common critical temperature $T_C^{local}$. The character of the  $\Delta_L(T)$ temperature dependence differs from $\Delta_S(T)$; this becomes obvious from the normalized temperature dependence $\Delta_S(T)/\Delta_S(0) \times \Delta_L(0)$ shown by squares in Fig.~10. The different behavior, confirms therefore, that the peculiarities observed in the dynamic conductance spectra are related to two distinct SGS's, and two different superconducting condensates, respectively. For comparison, we show the temperature dependence of bulk resistivity of the corresponding sample with nominal $x < 0.08$ (open rhombs in Fig.~10). The set of $\rho(T)$ data obtained with the polycrystalline samples showed that $\rho(T_C)$ nearly four times exceeds $\rho^{single}(T_C)$. Thus, the absolute values of $\rho^{single}$ were roughly estimated by normalizing of raw $\rho(T)$ by a factor of 4 in Figs. 10,11.

\begin{figure}
\includegraphics[width=20pc]{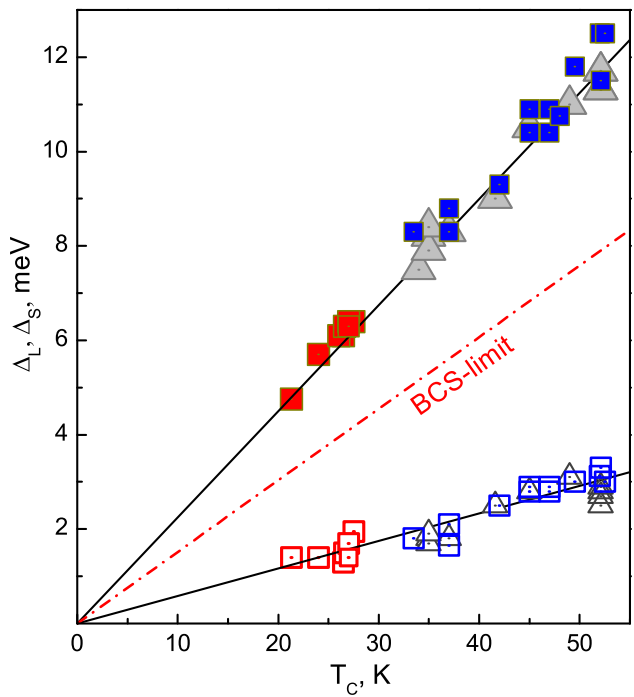}
\caption{The dependence of the large gap (solid squares) and the small gap (open squares) on the critical temperature for Sm-1111 with various thorium doping. The data of the present work are shown by squares (red squares depict data with nominal $x \lesssim 0.08$ series, blue squares \textemdash $x \approx 0.08 \textendash 0.3$). The data statistics obtained earlier by us with $x \approx 0.08 \textendash 0.3$ samples \cite{UFN,SmJETPL,FPSfit} is shown by triangles. BCS-limit 3.52 is shown by dash-dot line for comparison. Black lines are guidelines.}
\end{figure}

\subsection{Inter- and intraband coupling}
All temperature dependences of the large and small superconducting gaps we have measured agree well with predictions of two-band BCS-like Moskalenko and Suhl system of equations \cite{Mosk,Suhl} with a renormalized BCS-integral (RBCS) \cite{MgB2fit}. The equations describe the $\Delta_{L,S}(T)$ variation governed by diagonal (intraband) and off-diagonal (interband) coupling constants $\lambda_{ij} \equiv V_{ij}N_j$, where $N_j$ is the normal-state density of states at the Fermi level in the $j^{th}$ band, $V_{ij}$ is the matrix interaction elements ($V_{ij} \equiv V_{ji}$), $i,j = L,S$. To obtain theoretical $\Delta_{L,S}(T)$, we used the following fitting parameters: the relation between off-diagonal coupling constants $\lambda_{LS}/\lambda_{SL}$, the relation between intra- and interband coupling rate $\sqrt{V_LV_S}/V_{LS}$, and the eigen BCS-ratio for the small gap $2\Delta_S/k_BT_C^S$, here $T_C^S$ is the eigen critical temperature of the $\Delta_S$ condensate in a hypothetical case of the zero interband interactions $(V_{LS} = 0)$. Note, the sign of the interband $\lambda$ would not change their ratio, thus the sign can not be determined by such fitting procedure. The only restriction for these fitting parameters is obvious: $2\Delta_S/k_BT_C^S > 3.52$. The theoretical fits with 3 adjustable parameters are shown in Fig.~10 by solid lines; they capture correctly the experimental $\Delta_{L,S}(T)$ dependences.

\begin{figure}
\includegraphics[width=20pc]{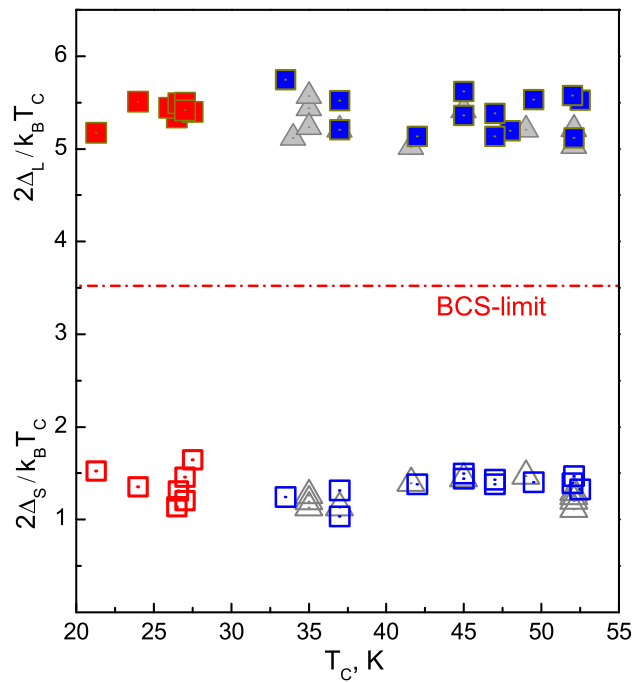}
\caption{The dependence of BCS-ratio for the large gap (solid squares) and for the small gap (open squares) on the critical temperature for Sm-1111 with various thorium doping. The data of the present work are shown by squares (red squares depict data with nominal $x \lesssim 0.08$ series, blue squares \textemdash $x \approx 0.08 \textendash 0.3$). The data statistics obtained earlier by us with $x \approx 0.08 \textendash 0.3$ samples \cite{UFN,SmJETPL,FPSfit} is shown by triangles. BCS-limit 3.52 is shown by dash-dot line for comparison.}
\end{figure}

In order to explore whether the generic $\Delta_{L,S}(T)$ temperature behavior is intrinsic to Sm-1111 compounds with various doping, we plotted in Fig.~11 several temperature dependences of both gaps obtained with Sm-1111 samples with various doping level. The $\Delta_L(T)$ dependences are presented by solid symbols, the $\Delta_S(T)$ \textemdash by open symbols. The temperature dependence of bulk resistivity near the superconducting transition for optimal single crystal is shown by gray open circles. Significantly, the $\Delta_L(T)$ dependences with $T_C^{local} = 26 \pm 1$\,K (solid circles, rhombs, and down triangles in Fig. 11) obtained with one and the same sample $\sharp 21$ look similarly. The value $\Delta_L(4.2\rm{K}) \approx 6.3$\,meV and the shape of its temperature dependence are reproducible and independent on both the contact resistance and the number of SnS-junctions in the array. Generally speaking, regardless of thorium doping, the typical features of $\Delta_{L,S}$ remain the same within all the $T_C$ range from 27\,K to 49\,K. The large gap temperature dependence passes slightly below the single-gap BCS-type (shown by dashed lines in Fig. 11), whereas the small gap dependence follows BCS-type only at $T \lesssim T_C^S$, then slow fades till $T_C^{local}$. The gap temperature dependences of the same type were observed in other oxypnictide groups, such as Gd-1111 and La-1111 \cite{UFN,FPSfit}. The observed $\Delta_{L,S}(T)$ behaviour is typical for a relatively weak interband coupling as compared to intraband one, and a higher normal density of states in the bands with the small gap.

In the presence of Coulomb repulsion between the quasiparticles the effective coupling constant should be calculated as $\lambda = \lambda^0 - \mu^{\ast}$ (here $\lambda^0$ is a full electron-boson coupling constant, and $\mu^{\ast}$ is a Coulomb repulsion). It is known, the experimental $\Delta_{L,S}(T)$ dependences are determined by namely effective coupling constant $\lambda_{ij}$ (see for example \cite{Kogan}), whereas the ratio of normal density of states (DOS) for both bands is determined by full coupling constant: $N_S/N_L = \lambda_{LS}^0 / \lambda_{SL}^0$. Supposing zero Coulomb repulsion as suggested in \cite{MazinRev,Mazin} for $s^{\pm}$ model ($T_C^{local} \approx 49$\,K), the relative coupling constants are $\lambda_L^0 : \lambda_S^0 : |\lambda_{LS}^0| : |\lambda_{SL}^0| = 1:0.65:0.3:0.03$, which leads to extremely high ratio of normal densities of states in the two bands $\lambda_{LS}/\lambda_{SL} = N_S/N_L \approx 10$. The latter is far from theoretical predictions, therefore one should use nonzero Coulomb repulsion constants $\mu_{LS}^{\ast}$ to estimate the full coupling constants $\lambda_{ij}$. In case of positive interband $\lambda_{LS}$, the DOS ratio is $N_S / N_L \approx 2$, and the relation between intra- and interband coupling rate approximately 2.5. The estimated relative $\lambda_{ij}$ are close to those calculated by us earlier in 1111-oxypnictides based on Gd, Sm, and La \cite{UFN}.

\subsection{Summary of the data}

By summarizing the gap values determined by IMARE spectroscopy of the Sm-1111 samples with $T_C = 21 \textendash 54$\,K, one may unreveal the influence of thorium doping on the superconducting properties (Figs.~12, 13). The $\Delta_L$ values are shown by solid symbols, the $\Delta_S$ values \textemdash by open symbols. The data of the present work are shown by squares. Blue squares correspond to samples with the nominal Th concentrations $x \approx 0.08 \textendash 0.3$ and well-reproduce the data obtained by us earlier (triangles) \cite{UFN,SmJETPL,FPSfit}. The pioneer data with nominal $x \lesssim 0.08$ series are shown by red squares, obviously, they follow the general course. Both superconducting gaps are in direct ratio with critical temperature as demonstrated in Fig.~12. Evidently, although the gap values are determined in the Andreev arrays with various cross-section, number of sequential contacts and, correspondingly, resistance of various Sm-1111 samples, the $\Delta_{L,S}(T_C)$ data in Fig.~12 is scattered insignificantly. We observe a good scaling of both superconducting gaps with critical temperature within the wide range of thorium doping and the wide range of critical temperatures, $21\,\rm{K} \leq T_C \leq 54\,\rm{K}$. The family of 1111-superconductors with Gd, La, and Ce, as well as FeSe chalcogenide also follow this tendency \cite{UFN}.

The linear $\Delta_{L,S}(T_C)$ dependences correspond to nearly constant BCS-ratios $2\Delta_{L,S}/k_BT_C$ for both gaps (Fig. 13). For the large gap, the BCS-ratio lies in the range $2\Delta_L/k_BT_C^{local} = 5.0 \textendash 5.7$. It is obvious that the interband interaction increases this ratio due to decreasing $T_C^{local}$. From fitting the $\Delta_{L,S}(T)$ dependences in the framework of Moskalenko and Suhl equations we have estimated the eigen BCS-ratio for the large gap: $2\Delta_L/k_BT_C^L = 4.1 \textendash 4.6$. The latter value exceeds the weak-coupling BCS limit 3.52 and points to a strong electron-boson coupling. The value obtained is close to those determined for 1111 oxypnictides by PCAR spectroscopy \cite{Miyakawa,Tanaka,Samuely}, nuclear magnetic resonance \cite{Mukuda}, and scanning tunneling microscopy \cite{Noat}. The BCS-ratio for the small gap $2\Delta_S/k_BT_C^{local} = 1.1 \textendash 1.6$ lies well below the BCS-limit, obviously, because $T_C^{local} \gg T_C^S$. By contrast, the eigen BCS-ratio for the small gap estimated from Moskalenko and Suhl fits is $2\Delta_S/k_BT_C^S = 3.5 \textendash 4$ (see also \cite{UFN,SmJETPL,FPSfit}). In Sm(Th)-1111, thorium atoms are located in Sm(Th)O-spacers, do not affect superconducting FeAs blocks directly and act as charge donors. Therefore, one may conclude that (Sm,Th) substitution do not change significantly the mechanism of superconductivity in Sm$_{1-x}$Th$_x$OFeAs.

\section{Conclusions}

By using intrinsic multiple Andreev reflections effect (IMARE) spectroscopy, we explored evolution of the superconducting properties of  Sm$_{1-x}$Th$_x$OFeAs compound with thorium doping. We determined the two superconducting gap values $\Delta_{L,S}$ for Sm$_{1-x}$Th$_x$OFeAs samples in a wide range of critical temperatures $T_C = 21 \textendash 54$\,K. We observed a good scaling of both $\Delta_L$ and $\Delta_S$ with $T_C$ in the whole explored range of $T_C$. The BCS-ratio for the large gap $2\Delta_L/k_BT_C^{local} = 5.0 \textendash 5.7$ and its eigen BCS-ratio (in a hypothetical case of zero interband coupling) $2\Delta_L/k_BT_C^L = 4.1 \textendash 4.6$ exceed the BCS-limit 3.52, thus suggesting a strong electron-boson coupling. For the small gap, $2\Delta_S/k_BT_C^{local} = 1.1 \textendash 1.6 \ll 3.52$, whereas its eigen BCS-ratio  $2\Delta_S/k_BT_C^S = 3.5 \textendash 4.0$ (when $V_{LS}=0$). The determined temperature dependences of the superconducting gaps $\Delta_{L,S}(T)$ are reproducible within the studied $T_C$ range and are well described with the two-band Moskalenko and Suhl system equations with a renormalized BCS-integral (RBCS). According to our estimates, the interband coupling is weaker than the intraband one by a factor of $\approx 2.5$, and the Coulomb repulsion constants $\mu^{\ast}$ are not negligible. The thorium substitution does not significantly change the mechanism of superconductivity in Sm$_{1-x}$Th$_x$OFeAs, making Sm(Th)O-spacers of crystal structure to act as charge reservoirs.

%


\appendix
\section{Theoretical Estimation of dI/dV Spectra Fine Structure in Case of Multiple Andreev Reflections}
\begin{figure}
\includegraphics[width=20pc]{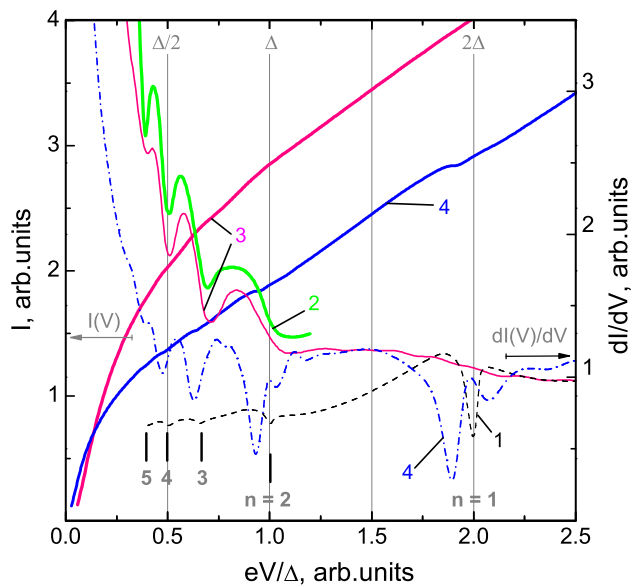}
\caption{Theoretical I(V) and dI(V)/dV of ballistic SnS contact obtained using various models: $1$ \textemdash OTBK \cite{OTBK}, $2$ \textemdash Arnold \cite{Arnold}, $3$ \textemdash Averin and Bardas \cite{Averin,Cuevas}, $4$ \textemdash K\"{u}mmel et. al \cite{Kummel}. The contact resistance was taken as unity. The positions of Andreev subharmonics $V_n = 2\Delta/en$ are labelled as $n = 1, 2, \dots$. $R_N = 1$.
}
\end{figure}
Theoretical modelling of MARE in a real SnS interface is a challenging issue so far. Since no analytical form of I(V) and dI(V)/dV is found, several numerical models are available, all for ballistic SnS contact in conventional superconductor with isotropic gap. The results are summarized in Fig. 14. The earliest model considering MARE was elaborated by Octawio, Tinkham, Blonder, and Klapwijk (OTBK model) \cite{OTBK} (spectrum $1$ in Fig. 14, $T = 0$, barrier height $Z = 1$). OTBK qualitatively showed a presence of subharmonic gap structure (SGS) comprising a set of dips at $V_n = 2\Delta/en$, where $n = 1, 2,\dots $ caused by MARE. Moreover, OTBK demonstrated that this formula is true up to $T_C$ \cite{OTBK}. Later theoretical studies by Arnold \cite{Arnold}, Averin and Bardas \cite{Averin}, and Cuevas, Poenicke, et al. \cite{Cuevas} derived a pronounced excess quasiparticle current at low bias voltages (so called foot area) in I(V) of SnS-contact. The spectrum obtained by Arnold \cite{Arnold} ($2$ in Fig. 14) was obtained with $T = 0$ and transmission probability $T^2 = 0.83$. Tight-binding model by Averin and Bardas \cite{Averin} and calculations by Cuevas et al. (based on that) \cite{Cuevas} predict less intensive feature with $n = 1$ (dI/dV and I(V) curves $3$ in Fig. 14, barrier transparency 95\%,
$T = 0$), whereas higher-order subharmonics $n = 2,3,\dots$ are more pronounced; nevertheless, the position of all subharmonics follow the formula (1). The results by Arnold, Averin and Bardas, and Cuevas, Poenicke et al. for high-transparency ballistic SnS contact are well-consistent as regards both the shape of SGS features, and the exponential increase of dI(V)/dV at $V \rightarrow 0$ (foot).
The model by K\"{u}mmel et al. \cite{Kummel} considers the case of high-transparent SnS, typical for our break-junctions. It also accounts carrier mean free path to contact dimension ratio, $l/2a$, a presence of quasiparticle Andreev band near the gap edge, and the finite temperature. The I(V) \cite{Kummel}, and the corresponding numerically calculated spectrum correspond to the case of $l/2a = 5$ and $T = 0.8T_C$ ($4$ in Fig. 14). The presence of the Andreev band causes satellite dips beyond the Andreev subharmonics. In the real case, one has no chance to establish experimentally the real shape of DOS distribution in the vicinity of the Andreev band, or/and even its presence. Without details, note the Andreev subharmonics become less intensive with $n$ increase, and the number of observed SGS dips roughly corresponds to $l/2a$ ratio. Of the most importance is the result \cite{OTBK,Kummel} that the position of SGS dips still follow the $V_n = 2\Delta(T)/en$ proportion within all temperature range till $T_C$. At $T \neq 0$, the SGS features become smeared rather than shifted from the positions given by formula (1) \cite{OTBK,Kummel}. This is the reason why the conventional $\Gamma$ broadening parameter should not taken into account.

Indeed, there is no solid accordance between MARE models in amplitude of Andreev dips and its variation with $n$ increase, as well as in the shape. Nevertheless, all the available models agree in the presence of SGS comprising dynamic conductance dips and its temperature behavior till $T_C$.

\begin{acknowledgments}
We thank Ya.G. Ponomarev, T. Shiroka, A. Charnukha for fruitful discussions, and S.N. Tchesnokov for the excellent technical assistance. Our work was supported by Russian Science Foundation grant 16-12-10507. The research has been partly done using the research equipment of the Shared facility Center at P.N. Lebedev Physical Institute RAS.
\end{acknowledgments}


\begin{thebibliography}{}
\bibitem{remark_1} Would the chains of intergrain contacts be independent, due to the chain accidental formation and nonequivalent conductivity through intergrain links, the $dI(V)/dV$-spectra of several contacts constituted on one and the same cleft, would have    been irreproducible. Upon increasing such intergrain links number, the spectra peculiarities would have become more smeared, an opposed behaviour to that observed  experimentally \cite{EPL}.
\end{thebibliography}

\end{document}